# Oxygen vacancy engineering of TaO$_x$-based resistive memories by Zr doping for improved variability and synaptic behavior

João H. Quintino Palhares[1,2], Yann Beilliard[2,3,4], Fabien Alibart[2,3,5], Everton Bonturim[6], Daniel Z. de Florio[1], Fabio C. Fonseca[7], Dominique Drouin[2,3,4] and Andre S. Ferlauto[1]

[1]CECS, Federal University of ABC, Santo André 09210-580, SP, Brazil
[2]Institut Interdisciplinaire d'Innovation Technologique (3IT), Université de Sherbrooke, Sherbrooke J1K 0A5, Canada
[3]Laboratoire Nanotechnologies Nanosystèmes (LN2) – CNRS UMI-3463 – 3IT, CNRS, Sherbrooke J1K 0A5, Canada
[4]Institut Quantique (IQ), Université de Sherbrooke, Sherbrooke J1K 2R1, Canada
[5]Institute of Electronics, Microelectronics and Nanotechnology (IEMN), Université de Lille, 59650, Villeneuve d'Ascq, France
[6]Department of Chemistry, School of Engineering, Mackenzie Presbyterian University, 01302907, São Paulo, SP, Brazil
[7]Nuclear and Energy Research Institute, IPEN-CNEN, São Paulo, 05508-000, Brazil
Corresponding author: quij2205@usherbrooke.ca, andre.ferlauto@ufabc.edu.br

E-mail: dominique.drouin@usherbrooke.ca, andre.ferlauto@ufabc.edu.br

**Abstract**

Resistive switching (RS) devices are promising forms of non-volatile memory. However, one of the biggest challenges for RS memory applications is the device-to-device (D2D) variability, which is related to the intrinsic stochastic formation and configuration of oxygen vacancy (V$_O$) conductive filaments. In order to reduce the D2D variability, control over the formation and configuration of oxygen vacancies is paramount. In this study, we report on the Zr doping of TaO$_x$-based RS devices prepared by pulsed laser deposition as an efficient means of reducing the V$_O$ formation energy and increasing the confinement of conductive filaments, thus reducing D2D variability. Our findings were supported by X-ray photoelectron spectroscopy, spectroscopic ellipsometry and electronic transport analysis. Zr doped films showed increased V$_O$ concentration and more localized V$_O$s, due to the interaction with Zr. DC and pulse mode electrical characterization showed that the D2D variability was decreased by a factor of seven, the resistance window was doubled, and a more gradual and monotonic long-term potentiation/depression in pulse switching was achieved in forming-free Zr:TaO$_x$ devices, thus displaying promising performance for artificial synapse applications.

Keywords: TaO$_x$, Zr:TaO$_x$, Oxygen vacancy engineering, memristor, variability, multilevel switching.

## 1. Introduction

Resistive switching (RS) memories based on oxide thin films are a promising technology for next-generation high-density nonvolatile memory applications and the hardware implementation of artificial neural networks,[1–4] due to their impressive switching operation performance in terms of speed, endurance, scaling and multistate resistance modulation.[5–8] However, both cycle-to-cycle (C2C) and device-to-device (D2D) variability in the switching characteristics still hinders the widespread use of this technology, as these reduce the average performance and generate cost increases due to the need for complex overhead circuits.[5,9,10] RS is generated by the creation and dissolution of conductive channels within an oxide layer in a capacitor-like structure, as a result of the nanoionic movement of oxygen vacancies ($V_O$) under the application of high electric fields.[11,12] An initial operation step called a forming process is typically required in order to generate sufficient $V_O$s to allow for the formation of conductive filaments (CFs). The stochastic nature of the forming and switching steps leads to slightly different CF configurations for each device, resulting in D2D and C2C variability.[5,10,12–16] Understanding and control of the formation, movement and arrangement of $V_O$s in RS devices is therefore paramount.[13,17,18]

Various strategies have been explored for the creation of excess oxygen vacancies in a controlled manner, in order to mitigate forming-induced variability.[19–21] The most common and direct method of $V_O$ generation is intrinsic doping, i.e., the use of low oxygen partial pressure during fabrication of the oxide layer, using methods such as reactive sputtering or pulsed-laser deposition (PLD). This approach has been particularly successful for tantalum oxide films.[20–22] Intrinsic doping has certain limitations, however, as it is difficult to control, relies on external parameters and is subject to instabilities in the processing conditions.[17,23–25] A more reliable alternative is extrinsic doping, which consists of adding dopant atoms with a valence that is lower than the host metal cation; this leads to the formation of $V_O$s in order to maintain electroneutrality.[12,17,26,27] Extrinsic doping also helps to reduce the randomness of $V_O$ generation. Theoretical studies of $Ta_2O_5$ doped with Ti, Zr and Al have indeed shown that the formation of $V_O$s is favored near the dopant species.[26] In these cases, charged $V_O$s are compensated and attracted by p-type dopants, forming dopant-$V_O$ complexes. These complexes trap $V_O$s, and can interconnect to form CFs. A reduction in variability has been experimentally observed for extrinsically doped $TaO_x$[28] and $HfO_2$ based[29,30] devices. It has been claimed that implanted or embedded dopant elements (e.g., Gd, Al) in $HfO_2$ help to localize $V_O$ formation and lead to more confined CFs, resulting in significantly lower device variability. This interaction between the dopant species and $V_O$s can change the switching dynamics and switching voltage operation,[18] meaning that extrinsic doping can be leveraged to further optimize switching performance for specific applications, such as artificial synapses.

In this work, we report the fabrication of $TaO_x$-based memristors with reduced D2D variability and controllable pulse switching dynamics, thanks to doping of the active layer with Zr. For comparison purposes, Zr-doped (Zr:$TaO_x$) and pure $TaO_x$-based devices were prepared in an oxygen-poor atmosphere using PLD. Oxygen-poor conditions are used to suppress electroforming step and its effect on the active layer properties. The chemical compositions of both types of devices were investigated using X-ray photoelectron spectroscopy (XPS) analysis and spectroscopic ellipsometry (SE). RS investigations in DC and pulsed mode revealed that D2D variability was decreased by a factor of seven, the resistance window was doubled and more gradual and monotonic long-term potentiation and depression (LTP/LTD) in pulse switching was achieved in Zr:$TaO_x$ devices. By analyzing the device current density–electric field (J–E) curves using a hopping transport model, we demonstrate that Zr:$TaO_x$ devices have a higher $V_O$ concentration and a lower variability of this concentration between different devices.

## 2. Methods

Devices were fabricated on a silicon substrate covered with 200 nm of $SiO_2$ (Fig. 1(a)). Sputtering was used to deposit tungsten (W) on the bottom (BE) and top electrodes. The latter were patterned to give 80 μm round pads, using UV photolithography and lift-off. $TaO_x$ and Zr:$TaO_x$ layers were deposited at room temperature via PLD (with a TSST system) using a $Ta_2O_5$ target and a $Ta_2O_5$:20 mol% $ZrO_2$ target, respectively, at a target−substrate distance of 45 mm, with a laser fluence of 3 J/cm$^2$ and a repetition rate of 10 Hz under an oxygen flow of 2 sccm. Layer thickness was estimated to be 10 nm by using the growth rate as determined by ES measurements. Based on previous studies,22 we used an $O_2$ partial pressure of $2\times10^{-2}$ mbar to induce the formation of slightly sub-stoichiometric active layers. The generation of $V_O$s by extrinsic and intrinsic doping was investigated using XPS (KRATOS Axis UltraDLD), SE (JA Woollam, model M-2000) and J–E conduction transport analysis of the devices. The electrical properties of the RS devices under DC and pulsed conditions were carried out with a Keithley S4200 semiconductor parameter analyzer and a Keithley 4225-PMU module. For all measurements, the BE was grounded and the signals were applied to the top electrodes.

## 3. Results and discussions

The thin films were analyzed using XPS and SE. The ratios (O/M) between oxygen and the sum of the metal atoms (M=[Ta] for $TaO_x$ and M=[Ta]+[Zr] for Zr:$TaO_x$) was determined by quantification of the Ta 4f, Zr 3d and O 1s peaks shown in Figs. 1(b-c). Due to the low p$O_2$ during deposition, $TaO_x$ and Zr:$TaO_x$ films had O/M ratios of 2.45 and 2.25, respectively. This is lower than the value of O/M=2.5 corresponding to the oxygen concentration of the stoichiometric tantalum pentoxide ($Ta_2O_5$), thus indicating the presence of $V_O$.[20,21] This effect was also confirmed by the change in the XPS Ta 4f peak, which suggests the presence of Ta cations in lower valence states (Ta$^{4+}$ and Ta$^{3+}$) (Fig. 1(b)). The energy positions of the chemical components Ta$^{5+}$, Ta$^{3/4+}$ and O 2s were 26.3, 25.7 and 22 eV, respectively. For the Ta 4f components, the area ratio was fixed at 0.75 and the spin orbit splitting at 1.9 eV.[20,31] The presence of Ta in a lower oxidation state is expected to result from the partial reduction of tantalum oxide (intrinsic doping), as given by the following reaction in Kröger–Vink notation[32]

$$2Ta_{Ta}^x + O_O^x \rightarrow 2Ta'_{Ta} + V_O^{\bullet\bullet} + \tfrac{1}{2}O_{2(g)}. \quad (1)$$

As expected, the Zr-doped films are more sub-stoichiometric, since the presence of Zr$^{4+}$ cations induces V$_O$ formation according to the reaction.

$$2ZrO_2 \xrightarrow{Ta_2O_5} 2Zr'_{Ta} + V_O^{\bullet\bullet} + 4O_O^x. \quad (2)$$

Furthermore, as shown in Fig. 1(c), the O 1s peaks have an asymmetric line shape that can be deconvoluted into two components centered at 531.3 and 530.0 eV; these can be assigned to non-lattice (O$_{nl}$) and lattice oxygen, respectively. The non-lattice oxygen peak can be associated with the presence of V$_O$.[33,34] The fact that Zr:TaO$_x$ films contain a larger percentage of non-lattice oxygen (26 %) compared to TaO$_x$ films (20 %) suggests that Zr$^{4+}$ substitution indeed promotes V$_O$ formation. This is in agreement with results reported by Park et al.[35] for KNBO films doped with Cu$^{2+}$, and by Kim et al.[36] for Si-doped tantalum oxide-based memristors. The XPS spectrum for the Zr 3d peak was deconvoluted into Zr 3d$_{3/2}$ and Zr 3d$_{5/2}$, centered at 182.4 and 184.7 eV respectively,[37–39] corresponding to Zr$^{4+}$ states (Fig. 1(d)). The nominal relative concentration of Zr with respect to the concentration of Ta atoms [Zr]/[Ta] was determined as 8%.

Figure 1(e) shows the optical absorption spectra obtained by fitting the ellipsometry data using a Tauc-Lorentz analytical function, which is widely used to describe the optical properties of amorphous and polycrystalline semiconductors and dielectric films.[40,41] In this function, the imaginary part of the dielectric function is described by a Tauc expression for photon energies above the gap, and is forced to zero below the band gap. The overall interband transition in the higher range of photon energies analyzed here is described by a Lorentz oscillator. Although this function can be used to describe pure stoichiometric dielectric Ta$_2$O$_5$ films, absorption in the sub-gap region is observed for sub-stoichiometric films. This feature can be modeled using a broad Lorentz function. Sub-gap absorption has been observed previously in sub-stoichiometric TaO$_x$ films[42], and was associated with the presence of V$_O$s.[43–45] From Fig. 1(e), it can be seen that the optical responses of both TaO$_x$ and Zr:TaO$_x$ exhibit considerable sub-gap absorption, thus confirming the XPS results. It is interesting to note that the optical absorption intensity is similar for both samples, even though the Zr-doped films have a higher V$_O$ concentration. One possible explanation for this is that the dopant-V$_O$ interaction may suppress the optical activity of the V$_O$ due to changes in the charge.[18,44]

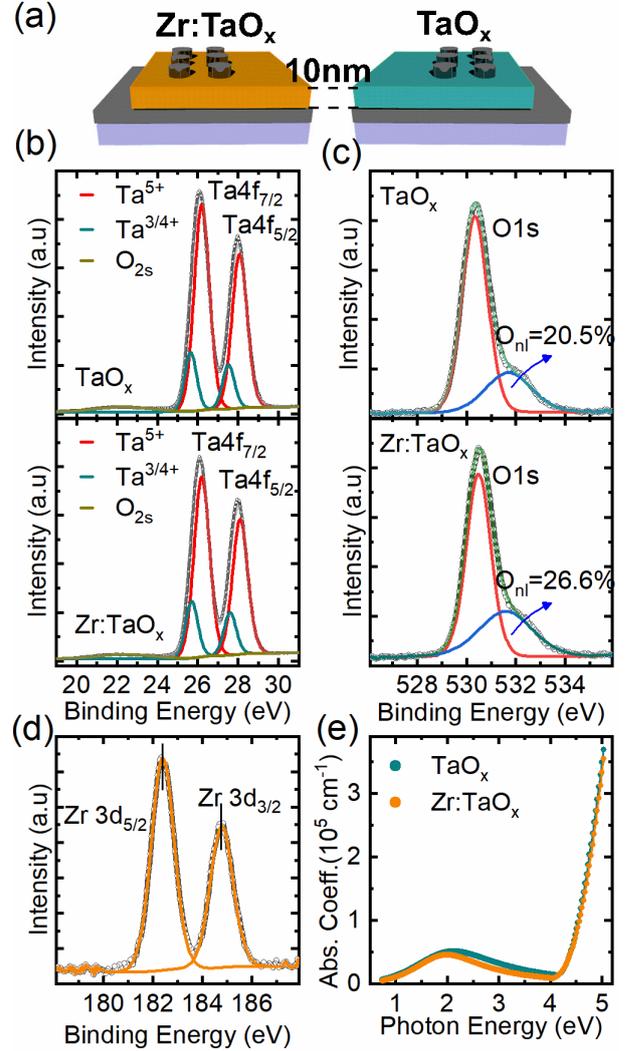

**Figure 1.** (a) Schematic of W/Zr:TaO$_x$/W and W/TaO$_x$/W RS devices and active layer characterization results: (b) XPS analysis of Ta 4f, (c) O 1s and (d) Zr 3d peaks for the Zr:TaO$_x$ and TaO$_x$ samples. (e) Absorption spectra of TaO$_x$ and Zr:TaO$_x$ obtained via SE.

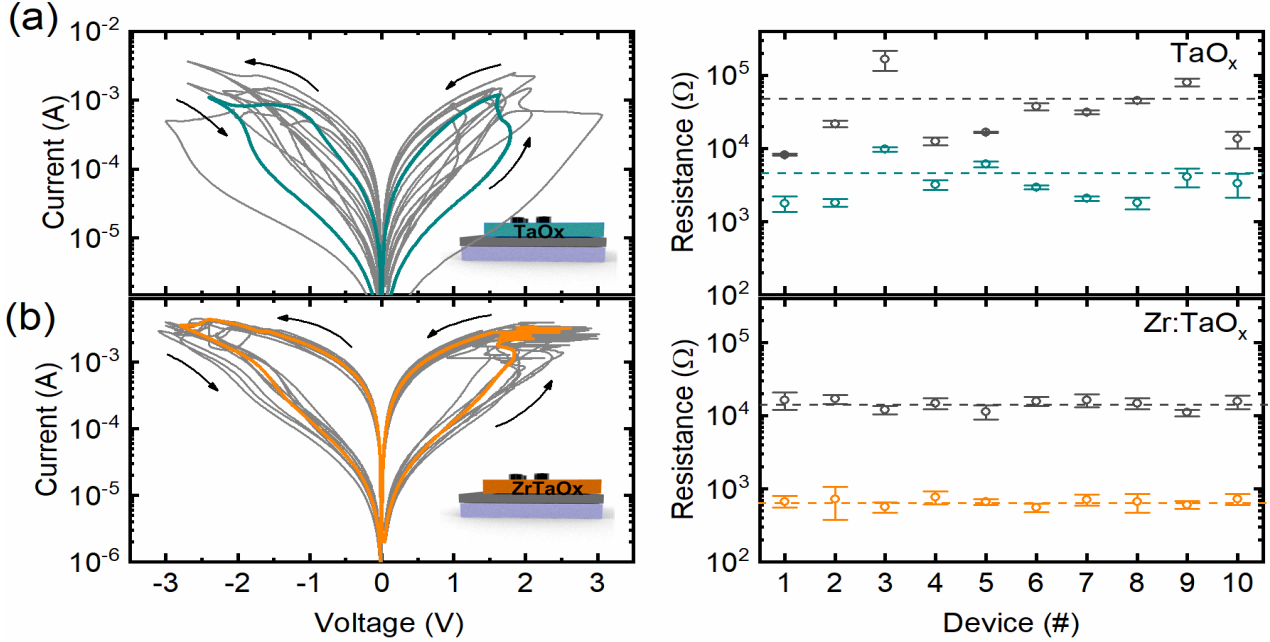

**Figure 2.** Current–voltage (I-V) characteristics for the 10 devices tested, with a statistical analysis of the resistance states per device: (a) W/TaO$_x$/W RS devices and (b) W/Zr:TaO$_x$/W RS devices. To improve visibility, the switching curve for one random device is shown in color. The resistance was measured at 0.2 V. The resistance values for the Zr:TaO$_x$ devices exhibit improved D2D variability.

To gain insight into the effects of Zr doping on the characteristics and variability of RS, current- and DC voltage-controlled sweeps were used for SET and RESET operations, respectively. For each device, in the SET and RESET operation, the current compliance and the maximum operation voltage values, respectively, were defined to maximize the device resistance window while avoiding irreversible damage. These values slightly vary among devices due to an intrinsic variation in the threshold voltage and maximum resistance window of each device. Figure 2 shows RS characterizations of W/TaO$_x$/W (Fig. 2(a)) and W/Zr:TaO$_x$/W (Fig. 2(b)) devices. Up to five full RS cycles were performed on each device, and the resistance state was measured at 0.2 V. For more RS cycles and performance information see supplementary material. In each RS plot, ten switching curves (the last RS cycle of each device) are given, corresponding to ten different devices with the same stack structure. The coefficient of variation (CV), defined as the standard deviation (σ) divided by the average value (μ) of a given device parameter, is used to account for the variability and dispersion. Table 1 summarizes the statistical data for main RS parameters. In the pristine state, most devices had resistance values ranging between the subsequent high resistance states (HRSs) and low resistance states (LRSs), and hence a forming process was not required. Few devices had initial resistance values higher than the HRS values, and these were switched with a voltage that was lower than in the subsequent SET and RESET operations. For both the Zr:TaO$_x$ and TaO$_x$ devices, the forming-free behavior was due to the high vacancy concentrations resulting from both intrinsic and extrinsic Zr$^{4+}$ doping.[20,21] The devices showed bipolar switching, with the SET operation occurring with a positive bias (V$_{SET}$) and the RESET operation with a negative bias (V$_{RESET}$) for all devices. The average values of V$_{RESET}$, V$_{SET}$ were slightly increased for Zr:TaO$_x$ devices (up to ~2.3 V as compared with ~2 V for TaO$_x$), an effect that can be ascribed to the dopant-V$_O$ interaction that may lead to defect clustering and consequent change in defect mobility. In addition, the current density values (@0.2 V) were slightly higher for the Zr:TaO$_x$ devices, which can be explained by the increased V$_O$ concentration. C2C variability in LRS and HRS resistance was similar for both the Zr:TaO$_x$ and TaO$_x$ devices, with average values for the CV of around 18% and 14%, respectively (see the curves in the supporting information). On the other hand, significant contrast was observed in the D2D variability, and especially in the resistance values in both the LRS and HRS states, in which the CV for the Zr:TaO$_x$ device in the LRS was reduced from 68% to 9% and in the HRS from 110% to 15% compared to TaO$_x$ (see Table 1). Another important feature observed for the Zr:TaO$_x$ devices was an increase in the average resistance window, from 12.5 for the TaO$_x$-based devices to 21.8 for the Zr-doped ones. A similar contrast was previously noted in a comparison of sub-stoichiometric to purely stoichiometric TaO$_x$ devices,[20] and was attributed to CF confinement and an increase in carrier concentration, which results in higher current densities. This localized Joule effect gives rise to greater filament dissolution and thus a larger resistance window.[20] These results indicate that Zr doping helps to improve RS performance by lowering D2D variability.

To further evaluate the effects of doping on the transport mechanisms, we conducted a carrier transport analysis by applying a hopping model that has been used previously in studies of tantalum oxide-based devices.[46–48] The J-E curves for the devices in both the pristine state and the HRS were fitted using the following expression[47,49,50]

$$J = qanv \, exp\left(\frac{qaE}{k_BT} - \frac{\phi_t}{k_BT}\right) \quad (3)$$

where a denotes the average distance between trap sites, $\phi_t$ is the trap energy, $k_B$ is the Boltzmann constant, q is the electronic charge, n is the carrier concentration and v is the attempt-to-escape frequency. As can be seen from Figs. 3(a)–(d), the I-V characteristics are well described by the hopping model (the fitted curves and parameters extracted for all devices are provided in the supplementary material). The values obtained for the average distance between traps for pristine Zr:TaO$_x$- and TaO$_x$-based devices were 1.0 ± 0.2 and 1.6 ± 0.7 nm, respectively (see Table 1). The HRS trap distances were 0.6 ± 0.1 and 0.7 ± 0.2 nm for Zr:TaO$_x$ and TaO$_x$ devices, respectively. The trap concentration for each type of device can be estimated by calculating[51,52]

$$[n_{Vo}] = a^{-3} \quad (4)$$

Pristine Zr:TaO$_x$ devices had a higher concentration of traps, with a value of (1.5 ± 1.4)×10$^{21}$ cm$^{-3}$ compared with (1.1 ± 1.8)×10$^{21}$ cm$^{-3}$ for TaO$_x$ devices. The same trend was observed for the HRS states: the Zr:TaO$_x$-based device has a trap concentration of traps (6.6 ± 3.4)×10$^{21}$ cm$^{-3}$ as compared to (3.6 ± 1.1)×10$^{21}$ cm$^{-3}$ for TaO$_x$. The transport mechanism in TaO$_x$ films depends critically on the V$_O$ concentration.

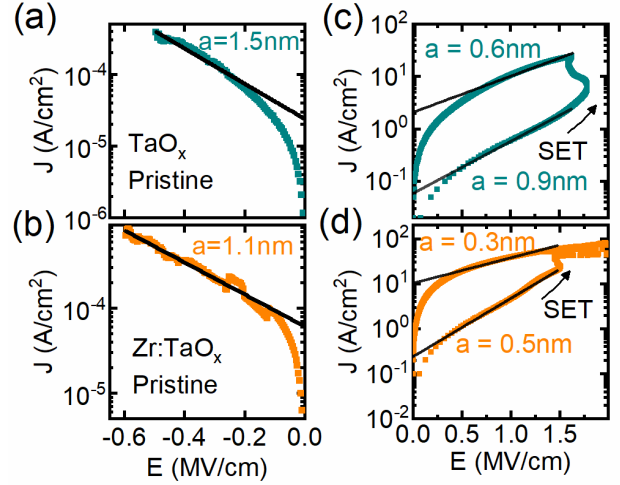

**Figure 3.** Experimental J-E plot and hopping model with fitted lines and extracted hopping distances for pristine (a) TaO$_x$ and (b) Zr:TaO$_x$ devices, and subsequent LRSs and HRSs for (c) TaO$_x$ and (d) Zr:TaO$_x$. Zr:TaO$_x$-based devices have a lower distance between traps in all resistance states.

TaO$_x$ exhibits Fermi glass behavior, in which V$_O$ complexes are trap sites and transport occurs via electron hopping.[53,54] An increase in V$_O$ concentration gives rise to a percolation process, and once the fraction of hopping sites exceeds a certain threshold, they form a continuous conductive path. The results in Fig. 3 indicate that the distances between traps for both Zr:TaO$_x$ and TaO$_x$ devices are in good agreement with reference values.[48,54–56] The shorter distance in the Zr:TaO$_x$-based devices explains the increased current density (@ 0.2 V). It is also important to emphasize that the variability of the devices is directly correlated with the variability in the distance between V$_{OS}$ or V$_O$ concentration, indicating that extrinsic doping is a more controllable way of changing the V$_O$ configuration.

**Table I:** Comparison of key parameters of Zr:TaO$_x$- and TaO$_x$-based memristors

|  | Zr:TaO$_x$ | | TaO$_x$ | |
|---|---|---|---|---|
|  | μ ± σ | CV (%) | μ ± σ | CV (%) |
| **LRS/HRS** | 21.8 ± 3.3 | 15.1 | 12.5 ± 7.6 | 60.8 |
| **R$_{LRS}$ (kΩ) @0.2V** | 0.70 ± 0.07 | 10.0 | 3.7 ± 2.7 | 73.0 |
| **R$_{HRS}$ (kΩ) @0.2V** | 14.5 ± 2.2 | 15.2 | 46.5 ± 49.6 | 106.7 |
| **J(A/cm$^2$) @0.2V – LRS** | 6.0 ± 0.70 | 11.6 | 1.4 ± 0.70 | 50.0 |
| **J(A/cm$^2$) @0.2V – HRS** | 0.30 ± 0.05 | 16.7 | 0.20 ± 0.10 | 50.0 |
| **V$_{SET}$ (V)** | 2.3 ± 0.40 | 17.4 | 1.7 ± 0.50 | 29.4 |
| **V$_{RESET}$ (V)** | 2.3 ± 0.50 | 21.7 | 2.0 ± 0.60 | 30.0 |
| **a (nm) – HRS** | 0.6 ± 0.1 | 16.7 | 0.7 ± 0.2 | 29.0 |
| **a (nm) – Pristine** | 1.0 ± 0.20 | 20.0 | 1.6 ± 0.70 | 43.8 |
| **[n$_{Vo}$] (10$^{21}$/cm$^{-3}$) – HRS** | 6.6 ± 3.4 | 51.5 | 5.5 ± 4.4 | 80.0 |
| **[n$_{Vo}$] (10$^{21}$/cm$^{-3}$) – Pristine** | 1.5 ± 1.4 | 93.3 | 1.1 ± 1.8 | 163.6 |

In order to assess the synaptic-like behavior of the devices, the switching dynamics were evaluated using pulsed measurements. We investigated comparative potentiation and depression characteristics using same time interval for LTP and LTD. As shown in Fig. 4(a), a sequence of 200 pulses for each LTD and LTP process was performed using time intervals of 200 µs and 600 ns, respectively. The LTD voltage amplitudes used were 2.4 V and 3.2 V, and LTP voltages were 2.1 V and 3.0 V for the TaO$_x$ and Zr:TaO$_x$ devices, respectively. The values of the applied voltage for the TaO$_x$ and Zr:TaO$_x$ devices were different because the voltage for RS operation is higher for the latter, as previously discussed. However, even at a higher voltage amplitude, it was observed that Zr:TaO$_x$ had more gradual potentiation and depression dynamics in its operation at a similar LRS/HRS ratio, as shown in the normalized pulse switching curves in Fig. 4(b-c). CV for both maximum and minimum conductance states are given for nine full LTD/LTP cycles. A non-monotonic increase in conductance can be seen in the first few potentiation pulses in the TaO$_x$-based device, and it has been claimed that this is related to a competition between thermal- and field-driven effects on the movement of V$_{OS}$ and the consequent CF stability.[57] This feature is detrimental for synaptic applications where weight update dynamics should be monotonic. The LTP curve for Zr:TaO$_x$ devices does not exhibit this problem however. This improvement can be attributed to the additional energy needed to overcome the attractive interaction between V$_{OS}$ and Zr cations. Although extrinsic doping generates a higher vacancy concentration, the trapping of V$_{OS}$ by dopant-Vo complexes also leads to a reduction in the overall V$_O$ kinetics. This tunability of the switching dynamics through extrinsic doping can be generalized to other p-type dopants, as previously mentioned by Lübben et al.[18]

Low C2C and D2D variability has been associated with a more reproducible formation and homogenous distribution of CFs.[14,58] In this work, the decrease in D2D variability cannot be attributed solely to the higher concentration of V$_O$, but is also related to the interaction between V$_{OS}$ and dopant species, which promotes V$_O$ localization around the dopant cations. This interaction has been well described for several doped oxides, in both experimental and theoretical simulation studies.[26–28] This localization may provide a guiding path for the formation of CFs, resulting in a less random distribution of CFs along the active layer. Moreover, simulation has shown that the presence of dopants reduces the V$_O$ formation energy.[26,27] It is interesting to note that Ambrogio et al.[59] have demonstrated that device variability is associated with fluctuations in V$_O$ energy barrier for hopping conduction. It can be argued that the presence of the dopant reduces this fluctuation, thus pinning the V$_O$ energy to a fixed value and reducing the variability. The V$_O$-dopant interaction is also reflected in the slower V$_O$ kinetics and the consequent change in the pulsed switching dynamics.

In this study, we have demonstrated that doping can be used to tune the V$_O$ content and consequently the defect trap density and conductive filament configuration in tantalum oxide-based memristor devices. It was shown that Zr doping (extrinsic doping) and control over the partial pressure of oxygen during PLD (intrinsic doping) can promote V$_O$ formation and lead to forming-free devices. In addition, extrinsic doping proved to be better in terms of controlling the $V_O$ configuration and content. The unique effect of Zr doping is that it localizes $V_O$s and consequently reduces the D2D variability. It was also used to tune the pulse switching dynamics for the devices. This enhancement in device parameters in terms of switching variability and dynamics reinforces the importance of exploring vacancy engineering for the tuning of memristor behavior for high-density memory applications and memristor-based artificial synapses.

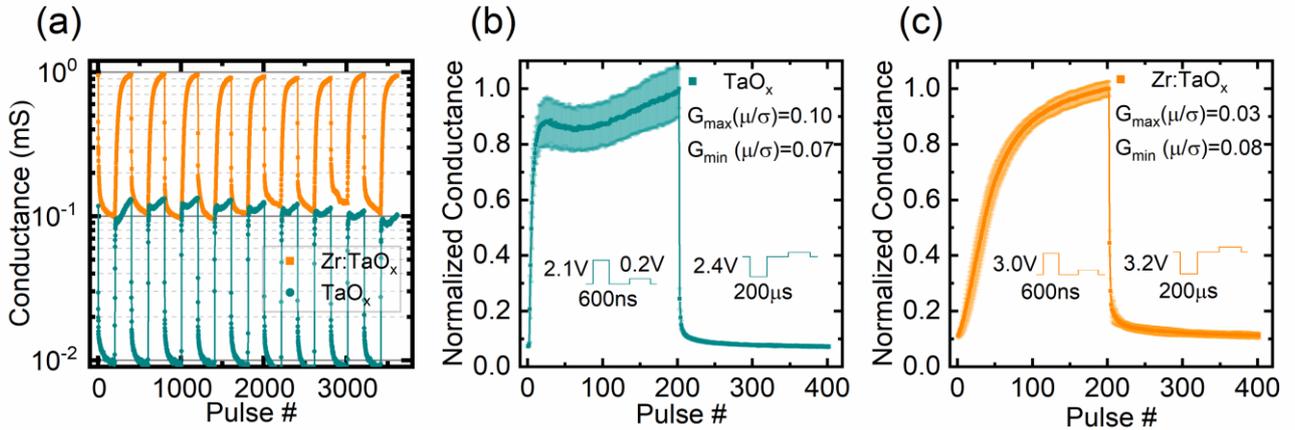

**Figure 4.** (a) Conductance values as a function of the number of pulses, plotted on a semi-log scale. Nine repeated pulsed-based LTD and LTP operations are shown, and synaptic weight emulation is demonstrated for the W/TaO$_x$/W (green) and W/Zr:TaO$_x$/W (orange) RS devices. (b, c) Average conductance and standard deviation as a function of the number of pulses for nine full LTD/LTP cycles of TaO$_x$ and Zr:TaO$_x$ devices. The latter exhibit a more gradual, reproducible, and monotonic variation of the conductance. CV of maximum and minimum conductance states are shown for both devices.

See the supplementary material for additional information on target preparation, C2C variability, device performance (endurance), hopping model fitted curves, retention measurements and thin film topography using an atomic force microscope.

**Acknowledgements**

This work was supported by Natural Sciences and Engineering Research Council of Canada (NSERC). Financial support of CNPq (INCT Carbon nanomaterials), CNEN, FAPESP (14/50279-4 and 17/11937-4) and Coordenação de Aperfeiçoamento de Pessoal de Nível Superior - Brasil (CAPES) - Finance Code 001, is acknowledged. We would like to acknowledge Abdelouadoud El Mesoudy and Wellington de Oliveira Avelino from 3IT for their help and insightful discussions regarding device electrical characterizations, and Sonia Blais from Plateforme de Recherche et d'Analyse des Matériaux (PRAM) at the Université de Sherbrooke for her support with XPS measurements. We also acknowledge LCPNano at the Federal University of Minas Gerais (UFMG) for providing access to the clean room facility and ellipsometry.

# Supplementary materials

# Oxygen vacancy engineering of TaO$_x$-based resistive memories by Zr doping for improved variability and synaptic behavior

João H. Quintino Palhares[1,2], Yann Beilliard[2,3,4], Fabien Alibart[2,3,5], Everton Bonturim[6], Daniel Z. de Florio[1], Fabio C. Fonseca[7], Dominique Drouin[2,3,4] and Andre S. Ferlauto[1].

[1]*CECS, Federal University of ABC, Santo André 09210-580, SP, Brazil*

[2]*Institut Interdisciplinaire d'Innovation Technologique (3IT), Université de Sherbrooke, Sherbrooke J1K 0A5, Canada*

[3]*Laboratoire Nanotechnologies Nanosystèmes (LN2) – CNRS UMI-3463 – 3IT, CNRS, Sherbrooke J1K 0A5, Canada*

[4]*Institut Quantique (IQ), Université de Sherbrooke, Sherbrooke J1K 2R1, Canada*

[5]*Institute of Electronics, Microelectronics and Nanotechnology (IEMN), Université de Lille, 59650, Villeneuve d'Ascq, France*

[6]*Department of Chemistry, School of Engineering, Mackenzie Presbyterian University, 01302907, São Paulo, SP, Brazil*

[7]*Nuclear and Energy Research Institute, IPEN-CNEN, São Paulo, 05508-000, Brazil*

**Contents:**

1. Target preparation and thin film deposition
2. Resistive switching and statistical data of all devices tested.
3. *I-V* sweep hopping model adjust of TaO$_x$ and Zr:TaO$_x$ devices.
4. Retention measurements.
5. Thin film surface topography measured by atomic force microscopy (AFM).
6. Performance: DC sweep resistance evolution and pulsed endurance.

## 1) Target preparation and thin film deposition

The TaO$_x$ and Zr:TaO$_x$ thin films were deposited by pulsed laser deposition (PLD, TSST; KrF excimer laser, 248 nm) operated at room temperature at a target−substrate distance of 45 mm, with a laser fluence of 3 J/cm$^2$ and 10 Hz repetition rate under an oxygen flow of 2 sccm at a pressure of 2×10$^{-2}$ mbar. The electrode and the active layer are deposited in different systems with time interval of few days. Both Zr:TaO$_x$ and TaO$_x$ samples are prepared at same conditions. The growth rate is 0.7 and 0.6 nm.min$^{-1}$ for TaOx and Zr:TaOx films, respectively. The PLD targets of Zr:TaO$_x$ were prepared by solid-state synthesis, mixing stoichiometric amounts of ZrO$_2$ (Optipure, Merck)) and Ta$_2$O$_5$ (Optipure, Merck). A mixture of 20% mol (0.3 g) of $ZrO_2$ and 80% mol (4.25 g) of $Ta_2O_5$ – 3%at Zr-$Ta_2O_5$ - were homogenously shacked for 15 minutes, pressed uniaxially at 120 MPa for 30 seconds and sintered at 1400 °C for 8 hours with heating and cooling rates of ±10 °C/min. The nominal relative Zr concentration with respect to the Ta atoms concentration [Zr]/[Ta] for the prepared target was ~12% and the one determined to our film was estimated to be 8% according to XPS analysis. Such deviation could be related to a variation in ablation yield and consequent non stoichiometric transfer during PLD process, as reported in previous works.[1,2] Phase purity was confirmed by XRD.

## 2) *I-V* sweep - Hopping model adjust.

Figure 1S and 2S show the *I-V* curves for all undoped $TaO_x$ and Zr:$TaO_x$ devices tested. Up to 5 full RS cycles were performed on each device and resistance is read at 0.2 V. All curves and its respective resistance standard deviation (σ) and average (µ) values are shown.

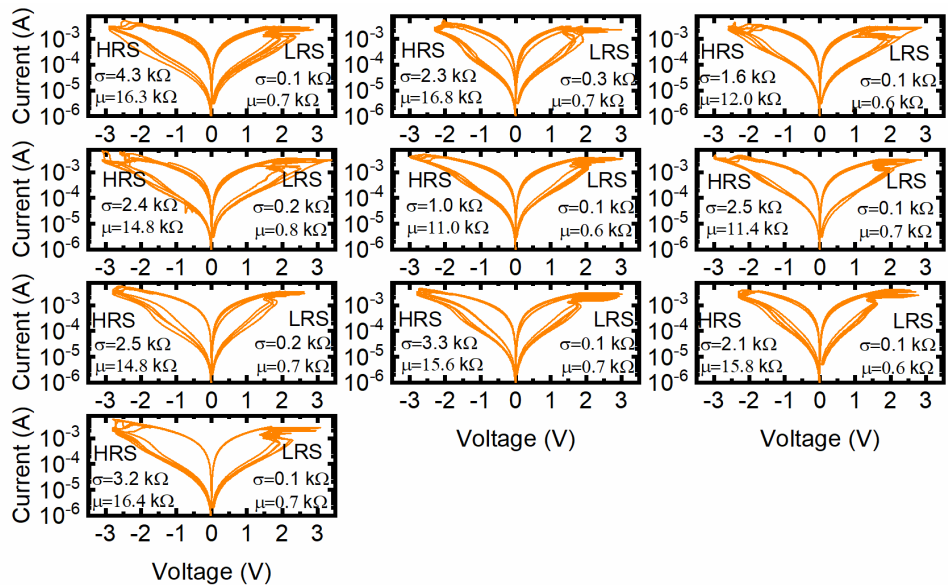

Figure 1S: the *I-V* characteristics for all Zr:$TaO_x$ devices tested and Its respective LRS and HRS resistance standard deviation (σ) and average (µ) values.

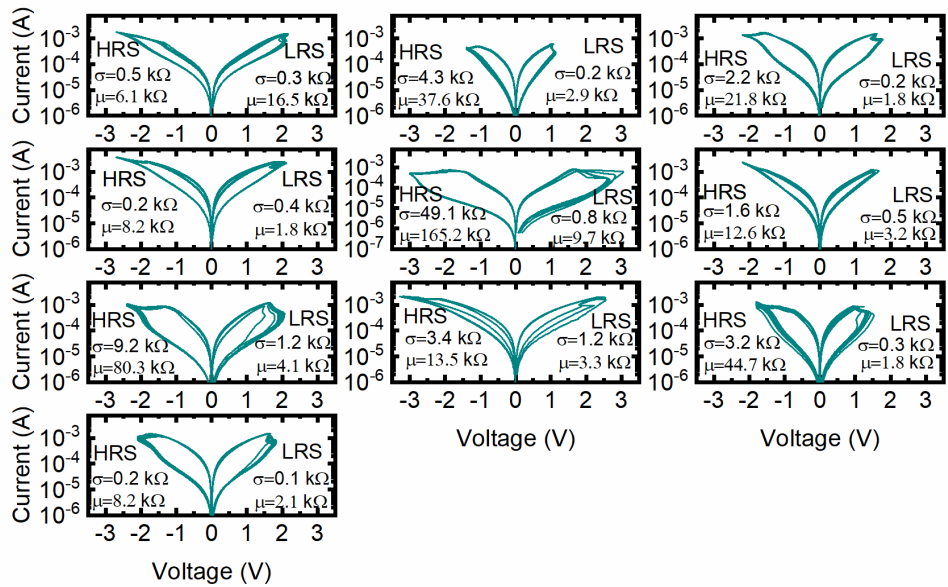

Figure 2S: the *I-V* characteristics for all TaO$_x$ devices tested and Its respective LRS and HRS resistance standard deviation (σ) and average (μ) values.

### 3) *I-V* sweep - Hopping model adjust.

Figure 3S and 4S show the *J-E* in SET operation for undoped TaO$_x$ and Zr:TaO$_x$ devices. The hopping model is used in HRS to estimate the distance between traps. As shown in Table 1 and in figures 1S and 2S, the mean distance between traps in Zr:TaO$_x$ devices and its standard deviation is reduce as compared to TaO$_x$ devices which is in accordance to the discussed effect of V$_O$ localization by Zr dopants.

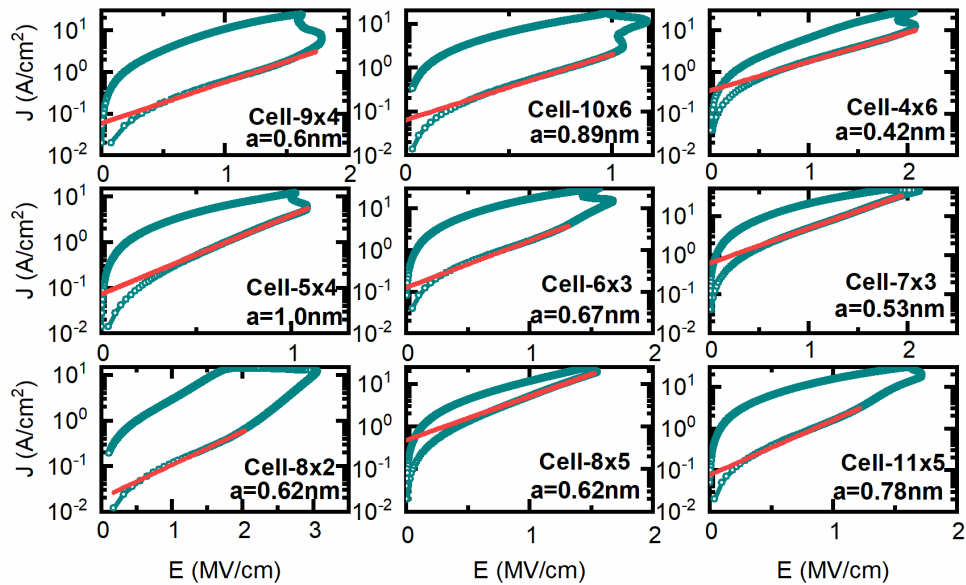

Figure 3S: *J-E* characteristics of W/TaO$_x$/W structure. The hopping model simulation curve for HRS is in red.

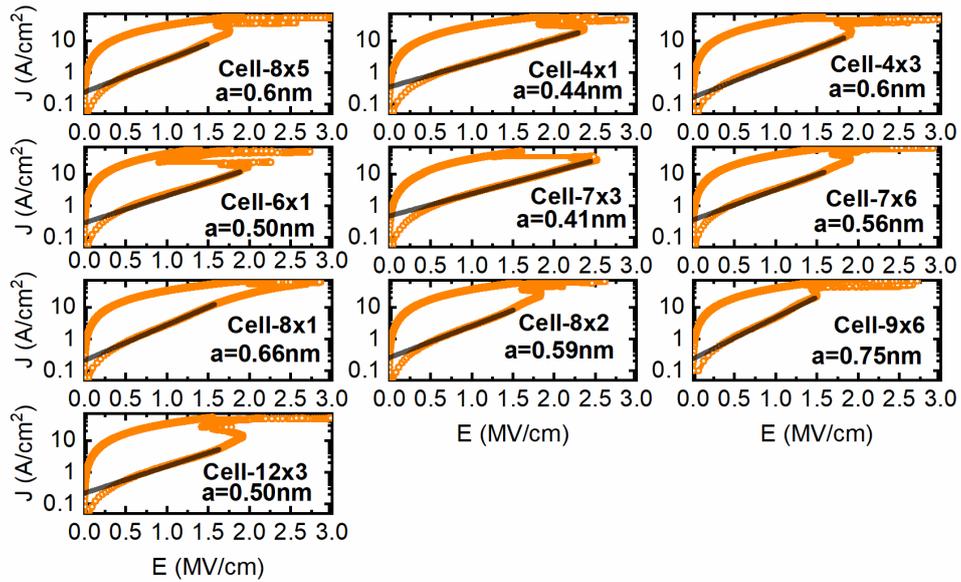

Figure 4S: *J-E* characteristics of W/Zr:TaO$_x$/W structure. The hopping model simulation curve for HRS is in black.

## 4) Retention measurements

Retention measurements were carried out for at least 2 hours for both Zr:TaO$_x$ and TaO$_x$ based devices. A reading was performed at 0.2V every 5 minutes. All resistive states stand for at least $10^4$ seconds, see figure 5S.

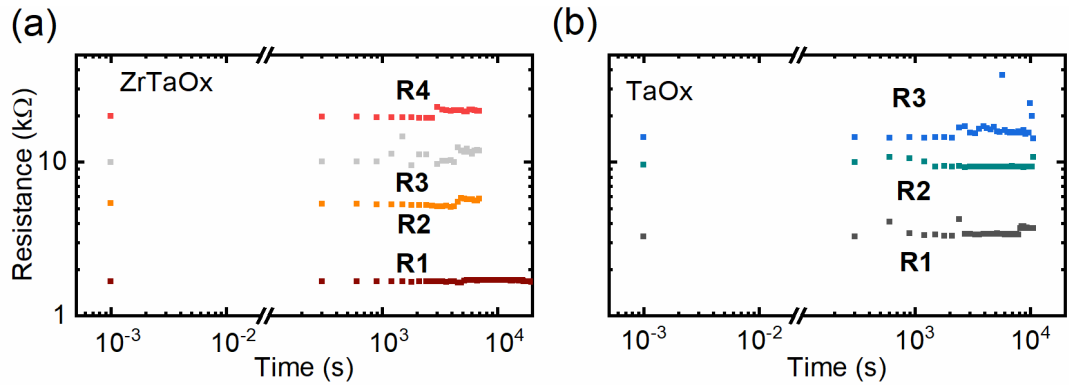

Figure 5S: Rentention measurements of (a) Zr:TaOx and (b) TaO$_x$ devices

## 5) Thin film surface analysis by AFM

Surface topography of thin films was acquired by using a Brucker MultiMode8 HR, the acquisition was done over an area of 0.25 µm$^2$. AFM measurements of TaO$_x$ and Zr:TaO$_x$ thin film surfaces reveal RMS roughness values of less than 0.4 nm. Such featureless surface indicates the amorphous nature of the oxides.

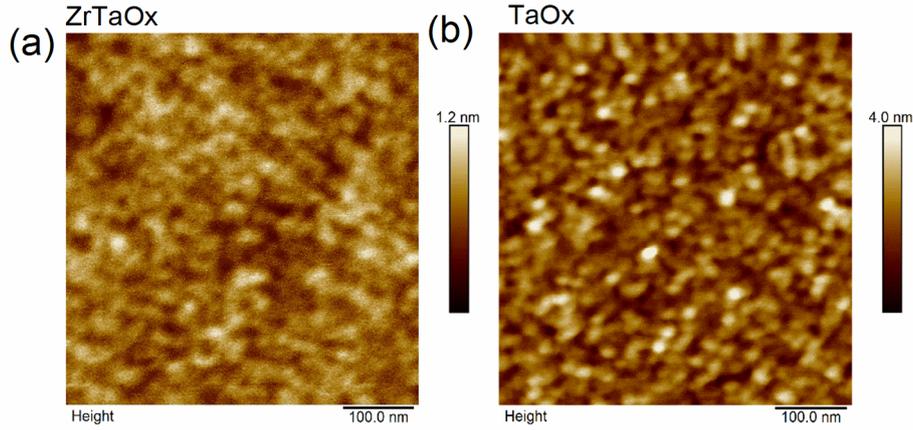

Figure 6S: AFM surface analysis of Zr:TaO$_x$ and TaO$_x$ thin films deposited by PLD.

**6) Performance: DC sweep resistance evolution and pulsed endurance:**

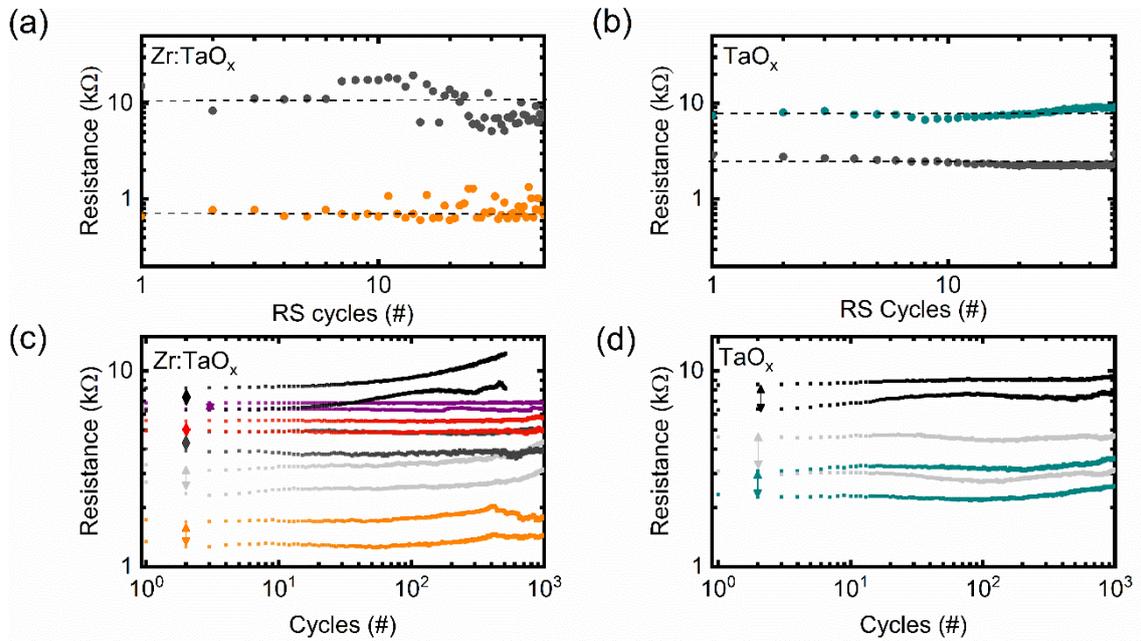

Figure 7S: Resistance (HRS/LRS) evolution of 50 consecutive full DC sweep cycles of a) Zr:TaO$_x$ and b) TaO$_x$ devices. c) Under alternate SET and RESET pulses endurance test of the Zr:TaO$_x$ and d) TaO$_x$ devices. $10^3$ switching cycles for different resistance levels and windows (represented with different colors) by varying SET and RESET pulse voltage amplitudes were obtained.

We have investigated the performance of Zr:TaOx and TaOx devices upon longer DC sweep resistive switching cycles (Fig. 7S(a-b)) and pulsed endurance (Fig. 7S(c-d)). The resistance evolution of Zr doped device exhibited a higher resistance window and a small degradation of high resistance state in DC sweep cycling compared to TaOx device, see figure 7S a) b). The average resistance values in both the HRS and LRS states for the Zr:TaO$_x$ device are 10.2 ± 3.0 and 0.7 ± 0.1 kΩ, respectively. On the other hand, the LRS and HRS states for the TaOx device are 8.2 ± 0.8 and 2.8 ± 0.3 kΩ. Besides degradation, The state instability in HRS shown by this Zr:TaOx device for 50 consecutive DC quasi-static RS cycles could be related to a quantum point contact behaviour that results from the confinement of the conductive filaments induced by doping. Similar instability was observed for TaOx devices by W. Yi et. al.[3] They found that device conductance starts to markedly vary near the first quantum of conductance, which corresponds to a resistance value of 12.9 kΩ. This value is close to the high resistance state of Zr:TaOx. Such resistance fluctuation is determined by a small number of bonds at the atomic scale quantum point contact (QPC) forming a hot spot in the CF. Nonetheless, such behaviour does not compromise its D2D variation as shown in figure 2 (b) since all devices are constrained to such resistance range between 10-20 kΩ that again coincides to the QC. In contrast, in pulsed endurance test (Fig. 7S(c-d)) both devices show up to $10^3$ switching cycles for different resistance levels and windows without observable degradation of device resistance. The average values of $V_{RESET}$, $V_{SET}$ are slightly higher for Zr:TaO$_x$ device, up to ~2.4 ± 0.4 V as compared with ~1.6 ± 0.1 V for TaO$_x$. As both devices present analog and incremental switching behavior different resistance windows and resistance levels are observed for different combinations of SET and RESET voltage amplitudes in pulsed endurance test. For both Zr:TaOx and TaOx devices in endurance test SET voltage amplitude varies from 2.6 to 3 V devices, and RESET voltage amplitude varies from -2.8 to -3.3 V. The pulse width varies from 150-200 µs. Resistance is read at 0.2 V.